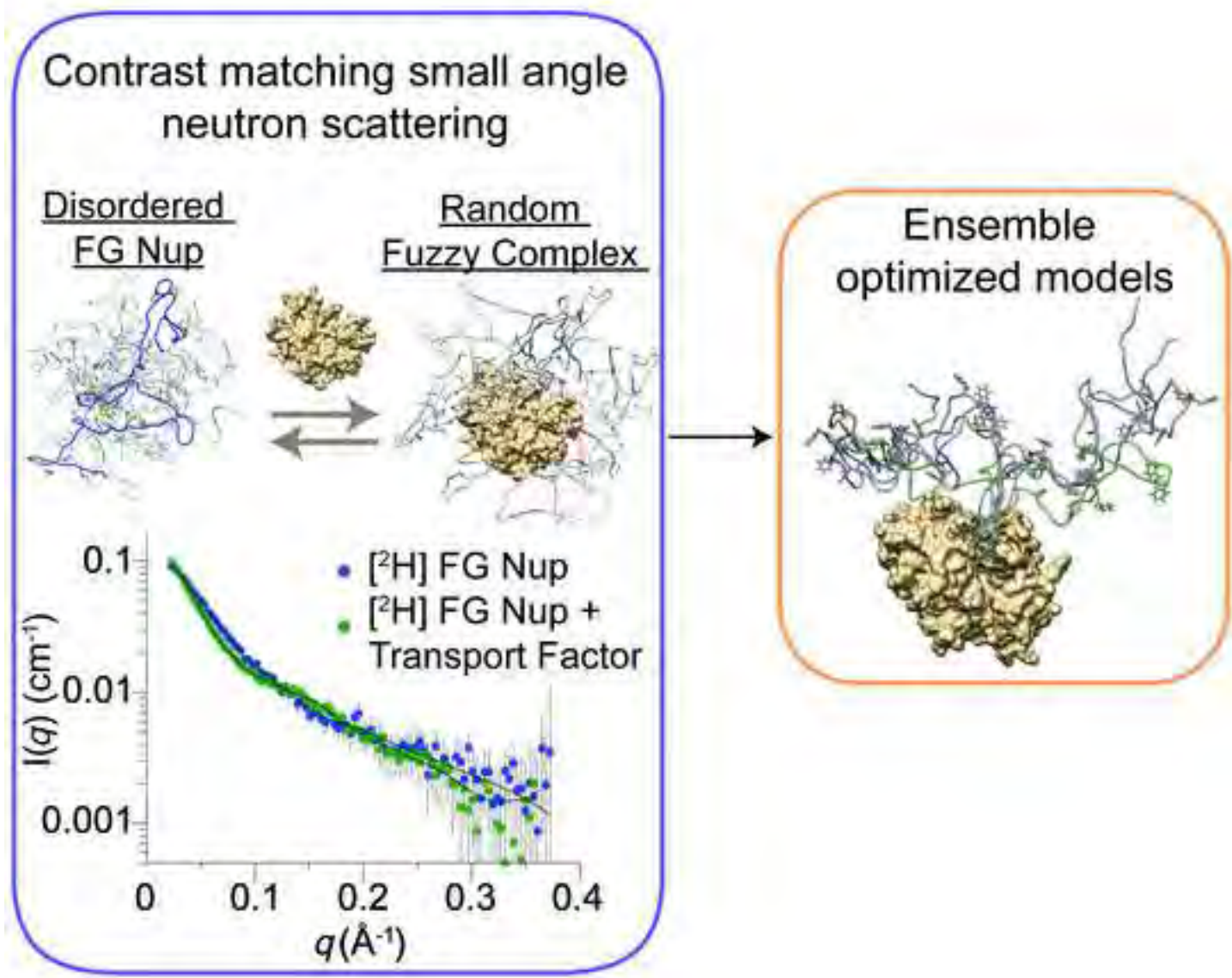

# Deciphering the 'fuzzy' interaction of FG nucleoporins and transport factors using SANS

Samuel Sparks[1], Deniz B. Temel[1,3], Michael P. Rout[2], and David Cowburn[1,4]*

[1] Departments of Biochemistry and of Physiology & Biophysics, Albert Einstein College of Medicine, Bronx, NY, USA

[2] Laboratory of Cellular and Structural Biology, Rockefeller University, New York, NY, USA

[3] Current address – Amgen Inc., Thousand Oaks, CA, USA

[4] Lead Contact

*Correspondence: david.cowburn@einstein.yu.edu

**SUMMARY**

The largely intrinsically disordered phenylalanine-glycine-rich nucleoporins (FG Nups) underline a selectivity mechanism, which enables the rapid translocation of transport factors (TFs) through the nuclear pore complexes (NPCs). Conflicting models of NPC transport have assumed that FG Nups undergo different conformational transitions upon interacting with TFs. To selectively characterize conformational changes in FG Nups induced by TFs we performed small-angle neutron scattering (SANS) with contrast matching. Conformational ensembles derived SANS data indicate an increase in the overall size of FG Nups is associated with TF interaction. Moreover, the organization of the FG motif in the interacting state is consistent with prior experimental analyses defining that FG motifs undergo conformational restriction upon interacting with TFs. These results provide structural insights into a highly dynamic interaction and illustrate how functional disorder imparts rapid and selective FG Nup – TF interactions.

**INTRODUCTION**

The selective permeability barrier of the nuclear pore complex (NPC) relies on a group of Phe-Gly rich nucleoporins (FG Nups) that contain large intrinsically disordered domains to generate an entropic barrier to nonspecific diffusion. Small molecules can freely diffuse through the NPC while larger macromolecules are impeded in a size-dependent manner (Knockenhauer and Schwartz, 2016; Timney et al., 2016). Larger macromolecules such as ribosomal subunits bypass the selectivity barrier by interaction with transport factors (TFs) which can permeate the disordered FG meshwork by virtue of making specific contacts with FG Nups. Exactly how FG Nup / TF interactions lead to rapid and selective transport is unresolved and several conflicting NPC transport models have been proposed (Grunwald et al., 2011; Schmidt and Gorlich, 2016). Despite significant effort (Beck and Hurt, 2017), there is no clear consensus of the underlying conformational preferences of FG Nups, the strengths of the FG Nup/TF interactions and any associated conformational changes which FG Nup undergo upon interactions with TFs. A molecular description, albeit reductionist, of these fundamental properties is required to comprehend the selectivity and rapidity of facilitated translocation of TFs through the NPC.

Recent studies utilizing nuclear magnetic resonance (NMR) and molecular dynamics (MD) simulations have revealed essential features that promote selective diffusion of TFs. Isolated FG Nups are highly dynamic, random coil polymers that remain disordered while engaged to TFs (Hough et al., 2015; Milles et al., 2015; Raveh et al., 2016). FG Nups interact with TFs using predominantly their FG motifs, and minimally their intervening spacer residues (Hough et al., 2015; Milles et al., 2015). By virtue of multiple TF interaction sites, FG Nups can make multivalent contacts with TFs, although whether or not high avidity interactions are critical to the NPC barrier function is still under debate (Hayama et al., 2017; Lim et al., 2015; Schmidt and Gorlich, 2016). FG Nups, represent a member of the unique class of intrinsically disordered proteins (IDPs) that form fuzzy interactions (Sharma et al., 2015; Wu and Fuxreiter, 2016). The interactions formed by FG Nups and TFs are consistent with the 'random' complex classification (Sharma et al., 2015) i.e. an ensemble of rapidly interconverting conformers, with multiple ligand sites (the FG motifs) dynamically contacting multiple TF-interaction sites without forming a stable secondary structure.

Prior NMR studies (Hough et al., 2015; Milles et al., 2015) have provided basic physicochemical properties of FG Nups and their interaction with TFs at high-resolution but in combination with small-angle scattering can provide a comprehensive description of the dynamic ensembles at different spatial and temporal scales (Hennig and Sattler, 2014). Indeed, this combination is particularly useful for characterizing IDPs (Receveur-Brechot and Durand, 2012). While sensitive to local dynamics and structural perturbations, NMR does not characterize global structural properties of IDPs. For example, the use of NMR data alone in ensemble modeling was not sufficient to differentiate between collapsed and extended conformers from disordered ensembles (Brookes and Head-Gordon, 2016). Therefore, it is unclear from our previous measurements (Hough et al., 2015) how TFs alter large-scale FG Nup dynamic structures. Previous reports, using different methods, offer conflicting observations of FG Nups undergoing (i) a reversible collapse upon interaction (Cardarelli et al., 2012; Lim et al., 2007; Ma et al., 2012), (ii) a moderate extension at high TF concentrations (Wagner et al., 2015), and (iii) no apparent global changes (Eisele et al., 2010; Milles et al., 2015; Zahn et al., 2016) upon interactions with TFs.

To resolve this issue, we extend our previous NMR measurements by utilizing small-angle neutron scattering (SANS) to capture prominent, potentially functional conformational features of FG Nups in an effort to specifically define the nature of these interactions beyond "fuzziness" (Figure 1A). SANS has unique advantages with respect to X-ray scattering for studying biological interactions as it can exploit the large difference in scattering length density of $^1H$ relative to $^2H$ nuclei to mask the signal from one type of component (Goldenberg and Argyle, 2014; Johansen et al., 2011). For example, with proteins in conjunction with deuterium labeling, SANS can selectively study the structural properties of a deuterated protein in protein complex by careful adjustment of the $H_2O/D_2O$ ratio of the solvent the such that it matches the scattering length density of the non-deuterated protein, eliminating their contribution to the scattering profiles (Heller, 2010; Stuhrmann, 2004). The contrast match point is known to occur at ~42% $D_2O$ for unlabeled proteins, whereas, partially deuterated proteins can be matched at near 100% $D_2O$ solvent, the latter of which is useful for reducing the incoherent background from $H_2O$ (Sugiyama et al., 2014). Scattering from IDPs are relatively weak with respect to the folded proteins (in this case, larger molecule weight TFs) such that small angle x-ray scattering (SAXS) of complexes of FG Nups and TFs are unlikely to resolve details of the mechanism of FG Nup interaction at intermediate length scales without accurate solvent matching using SANS with contrast matching. Here, we performed SANS with contrast matching to selectively observe FSFG-K, a previously defined model FG Nup construct, derived from the yeast Nsp1 and containing six equivalent FSFG repeat motifs (Hough et al., 2015) in the presence of two TFs, NTF2 and Kap95. Any differences in the observed scattering pattern of FSFG-K, with respect to the free state, can be solely attributed to changes in ensemble-averaged conformation induced by TFs. To our knowledge, this is the first study utilizing SANS with contrast matching to characterize an IDP involved in a 'fuzzy' IDP complex.

**RESULTS AND DISCUSSION**

We initially characterized the solution state of free FSFG-K. SANS profiles from [$^2H$]-FSFG-K, at 42% $D_2O$, and natural abundance FSFG-K, at 92% $D_2O$, display a featureless scattering curves with a power-law decay at $q > 0.05$ Å, typically observed for disordered proteins (Figure 1B) (Cordeiro et al., 2017). SANS measurements of a dilutions series of FSFG-K, the corresponding linear Guinier plots, and molecular mass estimates derived from the forward scattering, $I_0$, indicate the lack of significant inter-particle interference or aggregation (Figure S1A, S1B, and Table 1) (Trewhella et al., 2017). The radius of gyration, $R_g$ was determined by fitting the scattering data to Debye's law (Debye, 1947) which is valid over a larger $q$ range relative to standard Guinier analysis (Receveur-Brechot and Durand, 2012) (see STAR Methods for a discussion on extracting $R_g$ values from IDPs by small angle scattering). The Debye analysis yielded $R_g$ values of 36.2 ± 0.3 Å and 36.3 ± 1.0 Å (average value from the dilution series) for free [$^2$H]-FSFG-K (in 42% $D_2O$) and natural abundance FSFG-K (in 92% $D_2O$), respectively. These values are in good agreement with values obtained previously for FSFG-K from our MD simulation using the Tip4P-D water ($R_g$ 31.9 Å) (Raveh et al., 2016). Additionally, a dimensionless Kratky plot (Durand et al., 2010) displayed an initial increase, followed by a plateau at higher $q$ range, characteristic of an IDP (Figure 1B, inset). FSFG-K is then a fully disordered, random coil polymer, in full agreement with our previous NMR chemical shift analysis (Hough et al., 2015) and MD simulations (Raveh et al., 2016).

Contrast matched and inverse contrast matched experiments (Sugiyama et al., 2014) were performed i) at 42% $D_2O$, to match natural abundance TFs (NTF2 and Kap95) observing [$^2$H]-FSFG-K, and ii) at 92% $D_2O$, to match partially deuterated Kap95 observing natural abundance FSFG-K. In both experiments, excellent suppression of scattering from the TFs was observed (Figure 2A and Figure S2A). Upon addition of TF, the scattering profiles of [$^2$H]-FSFG-K displayed reduced intensity at $q \approx 0.06$ Å$^{-1}$ and a shoulder at $q \approx 0.12$ Å$^{-1}$ (Figure 2B). The observed shoulder was less pronounced at lower TF concentration, indicating these features are dependent on a high population of FSFG-K in the interacting state (Figure S2B-S2D). Due to poor yield of [$^2$H]-Kap95 in minimal $D_2O$ media, inverse contrast matching experiments were performed at a low concentration of partially [$^2$H]-Kap95. The apparent shoulder was, thus, not as pronounced, but there was some reduced intensity at intermediate scattering angles (Figure S2E and S2F). Nevertheless, the presence of these features indicates there is a shift in the ensemble averaged conformations of FSFG-K in the interaction state.

The overall size of FSFG-K increased upon addition of TFs. Debye analysis showed a TF concentration-dependent increase in the apparent $R_g$ upon addition of either NTF2 or Kap95, increasing ~15-30% from 36.2 ± 0.3 Å to a maximum of 47.9 ± 0.6 Å and 41.8 ± 1.2 Å at the highest concentrations of NTF2 (1.2 mM) and Kap95 (0.5 mM), respectively (Table 1). The increasing $R_g$ of FSFG-K with increasing TF concentration indicates that the observed changes are dependent on the degree of bound FSFG-K. Computing the radial distribution function, $P(r)$, further confirmed changes in the $R_g$ upon addition of TF and showed good agreement with the values obtained by Debye analysis (Figure 2C and Table 1). $P(r)$ curves additionally showed an increase in the maximum dimensions, $D_{max}$, from 127.0 Å in the free state to 168.0 Å and 143.5 Å in the presence of the highest concentrations of NTF2 and Kap95, respectively (Figure 2C and Table 1). An increase in the forward scattering, $I_0$, with increasing NTF2 concentration was also observed (Figure 2B and Table 1). While the increase in $I_0$ could be interpreted as subtle aggregation, dynamic light scattering (DLS) measurements of similarly prepared samples, both of the free components and as their complexes, are devoid of aggregate species (note that FSFG-K is unlabeled in the DLS experiments) (Figure S2G-S2J). Furthermore, the radius of hydration, $R_h$, determined for FSFG-K in the presence of either TF indicated that the molecular stoichiometry of the interaction is 1:1 (in terms of the number of molecules participating in the interaction). This implies the increase in $R_G$ and $I_0$ at high TF concentrations is not due to FSFG-K interacting with multiple TFs. We, therefore, can alternatively interpret the increase in $I_0$ as a change in the partial specific volume, $\upsilon$, of FSFG-K in the interaction state. $I_0$ is related to $(\upsilon^*\Delta\rho)^2$, where $\Delta\rho$ is the scattering contrast, thus a ~12% increase in the volume of FSFG-K could account for the ~25% change in $I_0$ (i.e. the observed increase from 0.115 cm$^{-1}$ to 0.142 cm$^{-1}$ for [$^2$H]-FSFG-K and [$^2$H]-FSFG-K + NTF2 (1.2 mM), respectively). Taken together, SANS data indicates that FSFG-K adopts larger, more extended conformations in the interaction state and does not undergo collapse upon binding as observed for other IDPs interactions (Green et al., 2016) as well as the reversible collapse observed for FG Nups using other approaches (Cardarelli et al., 2012; Lim et al., 2007).

Interestingly, in presence of either TF, the $P(r)$ curves displayed two peaks at inter-nuclear distances of ~15 Å and ~50 Å (Figure 2C). These real-space interatomic distances reflect the reciprocal space features present in the scattering profile (Figure 2B). The peak at ~50 Å would be the expected distance for the $i$, $i_{+2}$ FSFG motif (separation of 34 amino acids, see STAR Methods) and the peak at ~15 Å may reflect an increased local persistence length, $L_p$, of the polymer chain due to the local restriction of the 'strongly interacting' FSFG motif (Raveh et al., 2016). An increased persistence length is supported by modeling using the worm-like chain (WLC) model (Sharp and Bloomfield, 1968) (see STAR Methods). The expected range for persistence length of a typical IDP is 9-11 Å, which corresponds to 2-3 amino acid residues that are locally rigid (Perez et al., 2001). For [$^2$H]-FSFG-K, the derived values for $L_p$ increased upon addition of TFs from 8.9 ± 2.3 Å to 16.7 ± 1.2 Å (corresponding to 4-5 amino acids that are locally rigid) for the sample with 1.2 mM NTF2 (Table S1). Although for many samples some fitting parameters required constraint (See STAR Methods), good fits to experimental data were obtained and the resulting $R_g$ calculated from the model were in excellent agreement with Debye and $P(r)$ analysis (Table 1 and Table S1).

Ensemble modeling was performed to quantitatively assess changes in the ensemble-averaged conformation of FSFG-K upon interaction with TFs. The ensemble optimization method (EOM), which uses a genetic algorithm, was used for the selection of an ensemble of conformers whose weighted average scattering curve best reproduces the experimental data (Bernado et al., 2007; Tria et al., 2015). The structures within the selected ensemble are interpreted as containing the most prominent 'features' within the actual sample (Cordeiro et al., 2017). EOM selects these representative structures from a large initial pool of possible conformers. Two starting pools were used for selection to experimental SANS data: (1) a "random pool" of 100,000 all-atom, random coil models produced by the program TraDES (Feldman and Hogue, 2000) and (2) an "MD pool" of ~19,000 conformers obtained directly from the coordinates of FSFG-K in the presence of NTF2 (NTF2 coordinates removed) from our previous MD simulation on ANTON (Shaw et al., 2009) with the TIP4P-D water model (Raveh et al., 2016). Using this "MD pool" enables direct comparison of our MD simulations to experimental SANS data for FSFG-K interacting with NTF2.

In general, excellent agreement between the selected ensemble (best-performing sub-ensemble) and the experimental data was observed (Figure 3A, and Table S2). Ensemble modeling using the contrast matching SANS data with the initial selection from the "random pool" produced 2-4 conformers within the selected ensembles with $\chi^2$ values less than 0.4 (Table S2). The distribution of $R_g$ derived from the conformers in optimized ensembles (ensembles after 1000 generations of the genetic algorithm over multiple independent runs) for the free state of both the [$^2$H]- and natural abundance FSFG-K was similar to the $R_g$ distribution of "random pool". This confirms that the ensemble size adopted by FSFG-K is in good agreement with the expected size of a typical random coil (Figure 3B, Figure S1C, and S1D). In the presence of either TFs, optimized ensembles selected from the "random pool" display a shift in the $R_g$ distributions toward larger values, as expected, as FSFG-K forms more extended conformations in the interaction state, in agreement with both $P(r)$ and Debye analysis (Figure 3B and Table S2). We next computed the weighted average Cα-Cα distance maps from the selected ensembles (Figure 3C and Figure S3). The maximum dimensions of FSFG-K observed from the averaged distance maps increased from 87.4 Å in free state to 137.4 Å and 141.0 Å for [$^2$H]-FSFG-K in the presence of Kap95 (0.5 mM) and NTF2 (1.2 mM), respectively, comparable to $D_{max}$ values computed from $P(r)$ (Figure 3C and Table S2). If the conformers within the selected ensembles accurately reflect the experimental data, the organization of these structures should result in interatomic distances that mimic the real space distance computed from $P(r)$ in Figure 2C. Indeed, comparing the Cα-Cα distances between every Cα atom within a given conformer, plotted as a histogram over the selected ensemble, with that of the $P(r)$, produces comparable profiles (Figure S4A). Therefore, EOM appears to have successfully generated ensembles whose conformations reproduce the real space distances in the actual ensemble.

In the presence of high concentration of either TF, the conformers of FSFG-K selected from the "random pool" appear to adopt a highly extended, brush-like morphology (Figure 3D and Figure S3). However, despite the extended conformations, the conformers selected from fits to the scattering data of [$^2$H]-FSFG-K with high concentration TFs appear to contain 'loops' or 'kink' features followed by an extended segment in the chain. The Cα-Cα distance map indicates these features as a clustering of approximately 15 Å at the position of the loop conformation. These interatomic distances were also present at high frequency in the $P(r)$ (Figure 2C). For example, when EOM fitting was performed on [$^2$H]-FSFG-K in the presence of NTF2 (0.6 mM), 'kinks' in the structure (indicated by arrows in Figure 3D, middle) generally occurred in the middle of the extended conformation, forming a loop of ~10-20 residues with the apex of the loop composed of 3-6 residues. However, as evident by the distances maps of the individual conformers from the sub-ensembles (Figure S3) these conformers have no selective pressure to occur at the same residue position during fitting so these 'kink' features can appear blurred on an average distance map. The individual distances maps and the structures of the selected models indicate that in the presence of either TF sharp turns are consistently present in the selected ensembles when fit to the data using higher TF concentration (Figure S3). Although the selected ensembles derived from the free [$^2$H]-FSFG-K show some similar 'kink' features, the most prominent conformer within the ensemble (comprising 60% of the ensemble) appears as a typical crumpled coil with a low relative $R_g$ (30.9 Å). However, we caution that the specific conformers that comprise the selected ensembles are likely degenerate, and the theoretical from different combinations of conformers could reproduce the experimental data. While it is challenging to place significance on the specific conformers, we attempted to validate selected ensembles by demonstrating that the SANS data of [$^2$H]-FSFG-K in the presence of the lowest concentrations of either TF, can be adequately fit by a combination of previously optimized models selected of the free FSFG-K and FSFG-K in the presence of the highest concentration of TF. The underlying assumption is that features within the scattering profile reflect conformations of the bound FSFG-K that increase with a greater proportion of the interacting state occurring at greater TF concentration. Indeed, one model from the free [$^2$H]-FSFG-K selected ensemble (comprising 35.2% of the ensemble) and three selected models from the [$^2$H]-FSFG-K with NTF2 (1.2 mM) could accurately reproduce the scattering profile of [$^2$H]-FSFG-K with NTF2 (0.3 mM) with a $\chi^2$ = 0.33 (Figure S4B and S4C). Similarly, for Kap95, one of the selected models from the free [$^2$H]-FSFG-K selected ensemble (comprising 18.3%) and two of the three models from the [$^2$H]-FSFG-K with Kap95 (0.5 mM) selected ensemble reproduces the data of [$^2$H]-FSFG-K with Kap95 (0.25 mM) with a $\chi^2$ = 0.26 (Figure S4B and S4C). Thus, while the specific models maybe are degenerate, the conformational features appear conserved in the selected ensembles from the data of [$^2$H]-FSFG-K in the presence of high TF concentrations and likely reflect functionally relevant conformations.

Strikingly, a similar feature was also observed the Cα-Cα distance map calculated from the structures selected from the "MD pool" which reproduced SANS data of [$^2$H]-FSFG-K in the presence of high concentrations of NTF2 (Figure 3C and Figures S4D). The performance of the EOM fit to the SANS data of [$^2$H]-FSFG-K and NTF2 (0.6 mM), as well as, the $R_g$ distribution of the optimized ensemble were nearly identical with respect to the two starting pools ("Random pool" $\chi^2$ = 0.15, ensemble average $R_g$ = 40.4 Å, "MD pool" $\chi^2$ = 0.19, ensemble averaged $R_g$ = 39.4 Å) (Figure 3 and Table S2). Interestingly, three of the four conformers that comprise the sub-ensemble that best-fit the SANS data of [$^2$H]-FSFG-K + NTF2 (0.6 mM) with a selection from the "MD pool" are each in a similar pose and were derived from frames at a similar point in the MD trajectory (Figure 3D). The other conformer, which is accounts for 30% of the ensemble (Figure 3D, model colored tan), also contains a 'loop' feature (Figure 3D). This conformer was derived early in the simulation (after 57 ns) where the third FSFG motif of FSFG-K is bound to NTF2 in a conformation nearly identical to the conformation adopted in the crystal structure of NTF2 N77Y and a small

FSFG peptide (PDB: 1GYB) (Bayliss et al., 2002). This is illustrated in Figure 3C, where the grey arrow indicates the position of the bound FSFG motif. The MD simulations began with the coordinates of third FSFG motif, with the residues that interact with NTF2, constrained to their crystallographic positions and released at the start of the simulation. The bound FSFG in the crystal structures with NTF2 N77Y (Bayliss et al., 2002) and Kap95 (Bayliss et al., 2000) form similar conformations with the latter structure, showing six residues (KPAFSF) forming an extended conformation, which was denoted as "β-strand-like". Furthermore, in both crystal structures, the Gly forms a tight turn configuration, directing the rest of the FG Nup chain away from contacting other regions of the TFs (Bayliss et al., 2002). Similar 'loop' features were also observed for each conformer in the selected ensemble for [$^2$H]-FSFG-K with 1.2 mM NTF2 (Figure S4D). Therefore, based on ensemble modeling, we propose that the features in the SANS profiles for FSFG-K interacting with TFs reflect the bound FSFG motif forming extended conformations, contributing to the increased persistence length observed and that the Gly of FG facilitates the formation of the 'loop' observed in the selected ensembles. The loop conformation may help to avoid steric clashes between residues downstream of the FSFG motif and TFs surface enabling the Phe motifs probe for interaction sites efficiently. Previous chemical shift analysis of residues from FSFG-K closely matched their random coil values indicating no propensity for this conformation (Hough et al., 2015). We, therefore, rule out a conformational selection mechanism and suggest that the bound state conformation is associated with a high entropy penalty, providing a basis for the weak per-FG motif affinity expected from studies of similar FG Nup constructs (Milles et al., 2015).

The conformations of FSFG-K within the selected ensembles also resulted in a high frequency of interatomic distance of approximately 50 Å from fitting the SANS data with high TF concentration (Figure 2C and Figure S4A). The structural origins are most clearly evident from the models selected based on fitting to [$^2$H]-FSFG-K and Kap95 (0.5 mM) where the crumpled regions arising from 'kinks' in the middle and end of an otherwise linear polymer chain results in elevated distances of approximately 50 Å between the 'kink' regions (Figure S3). There are six equivalent FSFG motifs present in our construct, evenly spaced by 15 linker residues and only the FSFG motifs are involved in the interaction based on NMR analysis (Hough et al., 2015). The ~50 Å interatomic distances could reflect two or more FSFG motifs, from the same FSFG construct, binding to a single TF, although SANS is likely unable to definitely resolve whether multivalent interactions are occurring, especially in a highly dynamic system. Such conformations are reminiscent of those adopted by in the crystal structure of Nup214 bound to CRM1 where multiple FG motifs from Nup214 are engaging CRM1 (Port et al., 2015). In the selected ensemble derived from MD simulation with fit to the NTF2 at high concentration, internuclear distances of ~50 Å can are observed been adjacent 'loops' features (Figure S4D).

Taking our recent data together, we suggest the interactions between FG Nups and TFs are a version of a multisite "encounter complex" (Kozakov et al., 2014; Xu et al., 2008). The initial encounter occurs where FSFG motifs search for their specific TFs interaction sites where they then transition to a more strongly bound interaction mode (Raveh et al., 2016). In the 'strongly bound' state, the FSFG motifs have reduced dynamic motion as evidenced by the elevated [15]N relaxation rates ($R_2$) of the FSFG motifs in the presence of either Kap95 or NTF2 (Hough et al., 2015; Raveh et al., 2016). Here, our model based on SANS data extends previous NMR measurement and suggests that interaction of TFs induces small changes of local rigidity and effective persistence length of the FSFG motifs (and nearby residues). The bound motif is proposed here to adopt a conformation similar to that observed in the X-ray crystal structure (i.e. Gly forms a tight turn distending away from TF). The interaction of FSFG-K and TFs results in an increase in overall size of FG Nup and higher population of extended conformations due to (*i*) the rigid configuration of the bound FSFG motif which restricts compaction, similar to how IDPs containing polyproline helices imparts increased persistence length and larger than expected $R_g$ and $D_{max}$ relative to an IDP of the same number of residues (Boze et al., 2010); (*ii*) the turn-feature adopted by the Gly of the FSFG motif facilities loop formation helping to position the rest of the FG Nup chain away from contacting with other regions of the TF, and (*iii*) the influence of steric hindrance of due to the presence of the TF occupying space. In a 'fuzzy' complex, any local rigidity would be entropically unfavorable and would restrict the formation of static and high avidity FG Nup / TF interactions (Raveh et al., 2016), although it remains to be seen whether other flavors of FG Nups (Patel et al., 2007; Yamada et al., 2010) have similar interaction mechanisms.

Characterizing the nature of fuzzy complexes such as interactions involving FG Nups represents a significant challenge to the current toolbox of structural biology. Detailed understanding requires novel approaches and integration of a range of experimental and computational methods. SANS with contrast matching is an underrepresented technique for characterizing "unstructured" biology and measurements performed here provide important guides to how IDP complexes can be described in more detail than 'fuzzy'. By combining large-scale ensemble analysis, and validation by ensemble testing from two different data sources, we demonstrate that even flexible 'fuzzy' complexes can be approached using SANS with contrast matching.

## SUPPLEMENTAL INFORMATION

Supplemental information includes four figures and two tables and can be found with this article online at http://dx.doi.org/10.1016/j.str??????

## AUTHOR CONTRIBUTIONS

SS and DC contributed the concept and analysis. SS, DT conducted scattering measurement. MPR and DC supervised sample preparation. SS and DC drafted the paper, and all authors reviewed and finalized the ms.

## ACKNOWLEDGMENTS

The authors thank Dr. Paul Dominic Olinares and Dr. Brian Chait for graciously providing mass spectrometry analysis and Drs. Barak Raveh, and Ryo Hayama, for their helpful discussion. Dr. Shuo Qian provided technical advice at ORNL. This work was supported by NIH GM117212(DC) and GM109824(MR). This research used resources at the High Flux Isotope Reactor, a DOE Office of Science User Facility operated by the Oak Ridge National Laboratory. Simulations used the Anton special-purpose supercomputer provided by the National Resource for Biomedical Supercomputing (NRBSC), the Pittsburgh Supercomputing Center (PSC), and the Biomedical Technology Research Center for Multiscale Modeling of Biological Systems (MMBioS) through Grant P41GM103712-S1 from the National Institutes of Health. D. E. Shaw Research generously made the Anton machine available. Simulations also used the XSEDE facilities supported by NSF ACI-105375.

---

**Declaration of Interests:** The authors declare no competing interests.

---

**Figure 1. Experimental SANS data from the fully disordered FSFG-K.**

(A) A depiction of a conformational ensemble of free FSFG-K and a 'fuzzy' complex of FSFG-K interacting with Kap95 (Kap95 structure PDB id: 3ND2).

(B) SANS profiles of the free [$^2$H]-FSFG-K in 42% $D_2O$ (blue), and natural abundance FSFG-K in 92% $D_2O$ at concentrations of 20.2 mg/mL (indigo), 10.6 mg/mL (green), 5.3 mg/mL (black) and 2.7 mg/mL (red). (Inset) The dimensionless Kratky plot of [$^2$H]-FSFG-K and 20.2 mg/mL FSFG-K. The solid line represents the expected profile of a Gaussian chain.

---

**Figure 2. Contrast matching SANS data of [$^2$H]-FSFG-K in 42% $D_2O$ buffer in the presence of transport factors.**

(A) Demonstration of the quality of contrast matching for NTF2 and Kap95 in 42% $D_2O$ buffer. The red dashed lines are the moving average of the buffer scattering plotted on top of each scattering profile. The scattering curves are vertically displaced for clarity.

(B) SANS with contrast matching. Scattering profiles of [$^2$H]-FSFG-K (0.6 mM) (blue), [$^2$H]-FSFG-K (0.6 mM) + Kap95 (0.5 mM) (red) and [$^2$H]-FSFG-K (0.6 mM) + NTF2 (1.2 mM) (green). The SANS data is truncated to q = 0.25 Å$^{-1}$. The solid black line shows the fit of the P(r) displayed in panel C. The scattering curve of [$^2$H]-FSFG-K (0.6 mM) + NTF2 (1.2 mM) is vertically displaced for clarity and the free [$^2$H]-FSFG-K is plotted twice for comparisons with the SANS data of [$^2$H]-FSFG-K in the presence of transport factors.

(C) Radial distribution function, P(r), computed by the program GNOM (Svergun, 1992) corresponding to samples in panel B.

---

**Figure 3. Ensemble analysis using EOM fitting to SANS with contrast matching data.**

(A) The fit of the selected ensemble (best-performing sub-ensemble) to the experimental SANS data and a plot of the residuals of the EOM fit. From top to bottom, [$^2$H]-FSFG-K (0.6 mM), blue, (TraDES pool), [$^2$H]-FSFG-K (0.6 mM) + NTF2 (0.6 mM), orange, (TraDES pool), [$^2$H]-FSFG-K (0.6 mM) + NTF2 (0.6 mM), green, (MD pool), [$^2$H]-FSFG-K (0.6 mM) + Kap95 (0.5 mM), red, (TraDES pool). The scattering curves are vertically displaced for clarity.

(B) Plots of the $R_g$ distribution derived optimized ensembles after 1,000 rounds of EOM compared with the initial pool of TraDES models (Black) for the SANS data indicated. See also Table S2.

(C) Weighted average Cα-Cα distance maps computed from the selected ensemble. The black arrow indicates apparent distances from the ensembles of approximately 15 Å. The grey arrow present in distance map based on a selection from the MD pool indicates the sequence position of the FG motif bound to NTF2 in the MD simulation.

(D) Conformers of the best-performing sub-ensemble whose average SANS scattering are shown as solid lines in (A). The relative fraction that each conformer contributed is indicated. Left, [$^2$H]-FSFG-K, middle [$^2$H]-FSFG-K + NTF2 (0.6 mM) (TraDES pool), (right) [$^2$H]-FSFG-K + NTF2 (0.6 mM) with selection from the MD pool. Note the His-tag was not included in the MD simulation. The black arrows indicate the conformational feature, which results in the 15 Å distance in the distance map in C. The black arrows point to the similar sequence number. The grey arrow shows the position of FG motif bound to NTF2 in the MD simulation.

**Table 1. Comparison of the $R_g$ and I$_0$ values computed from Guiner (Forster et al., 2010), Debye (Debye, 1947) , and  the radial distribution function, P(r) (Svergun, 1992)  analyses.**

The expected $I_0$'s were calculated assuming the molecular weight derived from the amino acid sequence (14135.2 Da), a partial specific volume of 0.724 cm$^3$ g$^{-1}$, and a scattering contrast of either 3.357 x 10$^{10}$ cm$^{-2}$ (assuming 70% deuteration level) at 42% D$_2$O or -2.404 x 10$^{10}$ cm$^{-2}$ at 92% D$_2$O. Additionally, the assumption that 100% of exchangeable H's are available to solvent was made. Contrast calculations were made using the web service MULCh (Whitten et al., 2008). See STAR Methods section for a discussion on the calculation of $R_g$ from scattering methods. We note the $R_g$ derived from the SANS data of FSFG-K [375 µM] with [$^2$H]-Kap95 [75 µM] likely disagrees with the values obtained from [$^2$H]-FSFG-K in 42% D$_2$O buffer due to a slight mismatch in the inverse contrast match at low $q$ (Figure S2A).

| Sample | Guinier analysis | | Debye analysis | | P(r) | | | |
|---|---|---|---|---|---|---|---|---|
| | $R_g$ (Å) ($q_{max}*R_g$) | I$_0$ (cm$^{-1}$) | $R_g$ (Å) | I$_0$ (cm$^{-1}$) | $R_g$ (Å) | D$_{max}$ (Å) | I$_0$ (cm$^{-1}$) | Expected I$_0$ (cm$^{-1}$) |
| [$^2$H]-FSFG-K [0.6 mM] (8.5 mg/ml) | 32.0 ± 1.4 (1.33) | 0.110 ± 0.004 | 36.2 ± 0.3 | 0.115 ± 0.001 | 35.8 ± 1.1 | 127.0 | 0.115 ± 0.003 | 0.118 |
| [$^2$H]-FSFG-K [0.6 mM] Kap95 [0.25 mM] | - | - | 39.6 ± 0.8 | 0.087 ± 0.002 | 42.3 ± 2.7 | 147.0 | 0.092 ± 0.005 | 0.118 |
| [$^2$H]-FSFG-K [0.6 mM] Kap95 [0.5 mM] | - | - | 41.8 ± 1.2 | 0.097 ± 0.003 | 44.2 ± 1.7 | 143.5 | 0.103 ± 0.005 | 0.118 |
| [$^2$H]-FSFG-K [0.6 mM] NTF2 [0.3 mM] | - | - | 41.2 ± 0.3 | 0.125 ± 0.001 | 39.8 ± 1.9 | 142.8 | 0.122 ± 0.005 | 0.118 |
| [$^2$H]-FSFG-K [0.6 mM] NTF2 [0.6 mM] | - | - | 45.6 ± 0.5 | 0.143 ± 0.002 | 42.2 ± 2.2 | 151.0 | 0.138 ± 0.006 | 0.118 |
| [$^2$H]-FSFG-K [0.6 mM] NTF2 [1.2 mM] | - | - | 47.9 ± 0.6 | 0.143 ± 0.002 | 46.4 ± 2.9 | 168.0 | 0.142 ± 0.007 | 0.118 |
| FSFG-K [375 µM] + [$^2$H]-Kap95 [75 µM] | - | - | 53.2 ± 1.0 | 0.042 ± 0.001 | 47.2 ± 1.3 | 160.0 | 0.039 ± 0.001 | 0.038 |
| FSFG-K (20.2 mg/mL) | 31.5 ± 0.8 (1.04) | 0.1097 ± 0.002 | 34.8 ± 0.2 | 0.117 ± 0.001 | 33.5 ± 0.6 | 122.5 | 0.116 ± 0.001 | 0.143 |
| FSFG-K (10.6 mg/mL) | 32.6 ± 1.4 (1.04) | 0.069 ± 0.001 | 36.6 ± 0.3 | 0.071 ± 0.001 | 36.2 ± 1.0 | 139.7 | 0.072 ± 0.001 | 0.071 |
| FSFG-K (5.3 mg/mL) | 34.3 ± 2.9 (0.98) | 0.038 ± 0.001 | 36.2 ± 0.4 | 0.038 ± 0.001 | 35.2 ± 1.1 | 129.5 | 0.038 ± 0.001 | 0.038 |
| FSFG-K (2.7 mg/mL) | 33.6 ± 2.9 (1.15) | 0.019 ± 0.001 | 37.5 ± 1.1 | 0.019 ± 0.001 | 36.4 ± 1.3 | 129.5 | 0.019 ± 0.001 | 0.019 |

# STAR ★ METHODS

**CONTACT FOR REAGENT AND RESOURCE SHARING**

Further information and requests for resources and reagents should be directed to and will be fulfilled by the lead contact, David Cowburn, david.cowburn@einstein.yu.edu

**METHOD DETAILS**

Protein Expression and Purification

FSFG-K was expressed and purified as described previously (Hough et al., 2015). Briefly, BL21 (DE3) gold cells (Agilent) containing the expression plasmid (pET24a) were grown to an $OD_{600}$ of 0.6-0.8 and induced with 1 mM IPTG for 3 h at 37 °C. To produce [$^2$H]-FSFG-K, samples were grown in M9 minimal media containing 99% $D_2O$ (Cambridge Isotopes Laboratories, Tewksbury MA) with natural abundance glucose and $NH_4Cl$ as the sole carbon and nitrogen source, respectively (Shekhtman el al., 2002). Cells were lysed under denaturing conditions (8 M urea) and purified over Talon resin in lysis buffer (20 mM HEPES, 150 mM KCl, 2 mM $MgCl_2$, pH 7.4) with a protease inhibitor cocktail tablet, and with additional AEBSF and pepstatin A. The column was successively washed with lysis buffer with 8 M urea and the protein was eluted in elution buffer (20 mM HEPES-KOH, pH 6.8, 150 mM KCl, and 250 mM imidazole) with no urea. The elution was concentrated by centrifugal concentrators with 3 kDa MWCO (EMD Millipore, MA), and gel filtered using a Superdex S-200 column. Yeast NTF2 was expressed using a pRSFDuet expression plasmid and purified in an identical manner to FSFG-K except that urea was absent from all the buffers. Kap95 was expressed and purified as previously described (Hough et al., 2015). Cleavage of the GST-tag was performed by incubating Kap95GST with thrombin overnight. The sample was passed through benzamidine sepharose (GE) to remove thrombin, followed by glutathione Sepharose 4B (GE) resin to remove free GST. After thrombin cleavage, Kap95 was further purified by gel filtration using a Superdex S-200 column. Partially deuterated Kap95 was produced by growing Kap95 in M9 minimal media containing 85% $D_2O$ with natural abundance glucose and $NH_4Cl$ as the sole carbon and nitrogen source, respectively. MALDI-TOF mass spectrometry was performed to determine the percent deuterium incorporation achieved by comparing the mass of the partially [$^2$H]-Kap95 to a sample prepared under identical conditions but grown in regular media ($H_2O$). The amount of deuterium incorporation (of non-exchangeable protons) was 62% and this value was used to calculate the scattering length density.

Contrast matching experiments

Samples were prepared by dialysis of separate stock solutions of [$^2$H]-FSFG-K, NTF2 and Kap95 into 42% $D_2O$ buffer (20 mM HEPES, 150 mM KCl, 2 mM $MgCl_2$, pH 6.8). FSFG-K and [$^2$H]-Kap95 was dialyzed, separately, in identical buffer (with additional 0.5 mM TCEP) but with 92% $D_2O$ to match [62% $^2$H]-Kap95. In both cases, the dialysate was used to measure buffer scattering background. The web service MULCH (Whitten et al., 2008) was used to determine the contrast match point by calculating the scattering length density of [62%-$^2$H]-Kap95 and the buffer at different percentages of $D_2O$. The neutron scattering length density computed for 92% $D_2O$ buffer was 5.79 x $10^{10}$ $cm^{-2}$ and 3.41 x $10^{10}$ $cm^{-2}$ for natural abundance FSFG-K. The scattering length density for 42% $D_2O$ buffer was 2.36 x $10^{10}$ $cm^{-2}$ and 6.15 x $10^{10}$ $cm^{-2}$ for [$^2$H]-FSFG-K.

Protein concentrations for FSFG-K and NTF2 were measured by BCA assay kit (ThermoScientific, MA) following the manufacturer's instructions. The concentration of Kap95 was determined by $OD_{280}$ based on standard amino acid content ($\varepsilon_{280}$ = 85,260 $M^{-1}$ $cm^{-1}$).

Small angle neutron scattering

SANS measurements were conducted at the Bio-SANS instrument at the High-Flux Isotope Reactor, Oak Ridge National Laboratory (Heller et al., 2014). Data measurements of the contrast match series obtained at 42% $D_2O$ were acquired with a sample-to-detector distance of 2.53 m providing a $q$ range of 0.021-0.397 $Å^{-1}$. Data measurements made for the inverse contrast matching series at 92% $D_2O$ were obtained using a dual-detector setup with sample-to-detector distances of 6 m for the main detector and 1.1 m for the wing detector covering a $q$ range of 0.007 to 0.95 Å$^{-1}$. The scattering vector $q$ is defined as $q = 4\pi\lambda^{-1}\sin(\theta)$, where $\lambda$, is the neutron wavelength (6 Å) with a wavelength spread, $\Delta\lambda$, of 0.15 set by a neutron velocity selector. All samples were measured in a 1 mm path length at 25 °C. Samples containing [$^2$H]-FSFG-K for the contrast match at 42% D$_2$O were run for 3.5 h at 2.53 m. Natural abundance FSFG-K samples in 92% D$_2$O at 20.2 mg/mL and at 10.6 mg/mL were acquired for 1 h, while samples at 5.3 mg/mL and 2.7 mg/mL were acquired for 3 h and 3.5 h, respectively. Inverse contrast match experiments with 5.3 mg/mL FSFG-K and partially-[$^2$H]-Kap95 (1:0.2 molar ratio) were performed for 4.5 h. The scattering intensity profiles were obtained by azimuthally averaged the neutron detector counts, which were normalized to incident beam monitor counts. Detector dark current and pixel sensitivity were used for sensitivity correction and scattering from the quartz cell was subtracted. The software program PRIMUS from the ATSAS suite (Franke et al., 2017) was used to perform data merging as well as perform solvent subtraction. The software program DATGNOM, also from the ATSAS suite, was used to automatically determine the maximum intramolecular distance $D_{max}$ and compute of the distance distribution function P(r) using the fitted value of $R_g$ from Debye analysis as the initial estimate for expected $R_g$.

Dynamic light scattering

Dynamic light scattering measurements were made on a Dynapro plate reader (Instruments, Santa Barbara, CA) at 298 K. Samples were centrifuged prior to experiment and were placed in a temperature-regulated cell at a temperature of 25.0 °C Experiments were run in a 384 well plate format with 10 acquisitions of 5 s were acquired for each sample. The Dynapro software, DYNAMICS version 7.1.0.25, was used to analyze autocorrelation profiles with regularization fitting.

**QUANTIFICATION AND STATISTICAL ANALYSIS**

Calculation of $R_g$

Guinier approximation, which is the standard method of determining $R_g$, extracts the $R_g$ from scattering data within the small $q$ regime. In this regime $I(q)$ relates to $exp[-q^2R_g^2/3]$ and thus in the limit $q \to 0$, the slope of the scattering data, transformed as $log[I(q)]$ versus $q^2$, allows for estimation of $R_g$. For the Guinier approximation an upper limit for the $q$ range depends on the particle shape and homogeneity. For a sphere of uniform scattering density, Guinier analysis is valid in the limit of $q_{max}R_g < 1.3$. However, for highly disordered systems, Guinier analysis is known to be restricted to very narrow $q$-range ($q_{max}R_g$ <0.7-1) since higher order terms within the Guinier approximation become significant when distances between scattering elements of a polymer chain become large (Borgia et al., 2016). Therefore, determining $R_g$ by Guinier analysis is an experimental issue where the region of validity is either limited to only a few data points, because of limitation in instrument configuration, or the scattering data is hidden by the beam-stop and is therefore inaccessible. Guinier plots and analysis was performed using the software program SCATTER (Forster et al., 2010). For the Guinier analysis of the contrast matching datasets the instrument configuration (a sample-to-detector distance of 2.53 m) did not allow for sufficient coverage of the low $q$ Guinier region to allow for analysis within the limit of $q_{max}R_g < \sim$1. The free [$^2$H]-FSFG-K in 42% D2O, which had a smallest relative $R_g$ was subsequently the only dataset to be fit albeit with only the first 5 data points. Guinier analysis of this dataset therefore should be considered an inaccurate measurement of the $R_g$.

The Debye equation (Debye, 1947), which describes the behavior of a Gaussian chain, has been suggested as an alternative as it can be applied to a much larger $q$ range ($q_{max}R_g < 3$) (Perez et al., 2001; Receveur-Brechot and Durand, 2012). Debye analysis was used to determine the $R_g$ by the equation:

$$\frac{I(q)}{I(0)} = \frac{2}{x^2}(x-1) + e^{-X} \text{ (eq. 1)}$$

where $= q^2R_g^2$. The $q$ range used in for fitting was optimized for each sample. However, in some cases it has been noted that the Debye model is known to fit poorly for chains with excluded volume at larger $q$ (Petrescu et al., 1998).

The radial distribution function, $P(r)$, is obtained by an indirect Fourier transform of the scattering pattern related by the equation (Svergun, 1992):

$$P(r) = \frac{r^2}{2\pi^2} \int_0^\infty \frac{q^2 I(q) \sin(qr)}{qr} dq \text{ (eq. 2)}$$

$P(r)$ is equal to zero at $r = 0$ and is expected to decay smoothly to zero at $r = D_{max}$. The $R_g$ can be calculated from the P(r) by the function:

$$R_g^2 = \frac{\int_0^{Dmax} r^2 P(r) \, dr}{2 \int_0^{Dmax} P(r) \, dr} \text{ (eq. 3)}$$

Calculation of $P(r)$ and resulting $R_g$ was performed using GNOM (Svergun, 1992). For disordered systems, $R_g$ values obtained from the $P(r)$ are considered a more reliable estimate relative to the valued obtained Guinier approximation (Perez et al., 2001). Durand and co-workers state that the $R_g$ calculated by Guinier analysis for a completely unfolded protein would yield values that are systematically smaller than the "true" $R_g$ whereas values calculated from Debye equation and $P(r)$ were considered more accurate. Our results are in agreement showing smaller $R_g$ values from fits to the Guinier approximation whereas values from Debye and $P(r)$ are in better agreement (Table 1). Accurate extraction of $R_g$ via $P(r)$ methods requires a minimum $q$ value of $\sim\pi/D_{max}$. There is, however, an inherent uncertainty deriving the $D_{max}$ and care is usually taken in deriving an optimal value (Trewhella et al., 2017). In some cases, underestimation of the maximum dimensions for an unfolded protein can lead to an underestimation of the $R_g$ (Borgia et al., 2016). In this study, the $R_g$'s calculated from the $P(r)$, the Debye function and the values obtained from ensemble modeling agree reasonably well ensuring confidence in the obtained values.

<u>Worm-like chain modeling</u>

The worm-like chain model (Kratky-Porod chain) is a form factor equation used to describe a polymer chain with a contour length, $L$, and a persistence length, $L_p$. The persistence length accounts for local rigidity of the random coil and is parameterized in the model as the statistical length, or Kuhn length, $b$, defined as twice the persistence length ($b = 2L_p$). The expression for the form factor used to describe the Kratky-Porod chain is written as (Sharp and Bloomfield, 1968):

$$\frac{I(q)}{I(0)} = \frac{2}{x^2}(x - 1 + e^{-x}) + \frac{b}{L}(\frac{4}{15} + \frac{7}{15x} - [\frac{11}{15} + \frac{7}{15x}]e^{-x}) \text{ (eq. 4)}$$

where $x = \frac{q^2 L b}{6}$. From the fitted values of $L$ and $b$, one can compute the radius of gyration $R_g$ by:

$$R_g^2 = b^2 \left[(\frac{y}{6} - \frac{1}{4} + \frac{1}{4y} - \frac{1}{8y^2})(1 - e^{-2y})\right] \text{ (eq. 5)}$$

where $y = \frac{L}{b}$. This function is valid for $\frac{L}{b} > 10$ and $q < \frac{3}{b}$. Note that for the samples with [²H]-FSFG + Kap95 at 42% $D_2O$ these conditions could not be maintained.

From the calculated $L_p$, the average number of amino acid residues that are locally rigid, $n_p$, can be calculated by $n_p = L_p/l_0$, where, $l_0 = 3.78$ Å, the distance between two adjacent Cα residues. Fitting was performed in the limit of $q < 0.13$ Å⁻¹. For many samples, the value of $L$ needed to be constrained to the theoretical limit. The maximum $L$, is defined as $L = nl_0f$, where $n$ is the number of amino acids ($n = 133$), $l_0 = 3.78$ Å, and $f$ is a geometric factor ($f = 0.95$). FSFG-K, therefore, has a maximum $L$ of 477.6 Å. As noted previously (Daughdrill et al., 2012; Perez et al., 2001), a mathematical limitation inherent to WLC model, the dependence of the function on the product $Lb$, can lead to high error in the fitted value $L$. With the imposed constraint, we were able to observe good fits to experimental data ($R^2 > 0.99$ for each sample with the exception of the two lowest concentration (5.3 and 2.7 mg/mL) apo FSFG-K samples in 92% $D_2O$).

From polymer theory, we estimate the root-mean-square distance, $\langle r^2 \rangle^{\frac{1}{2}}$ (in decimeters), between ends of an ideal chain undergoing random walk as:

$$\langle r^2 \rangle^{\frac{1}{2}} = a\sqrt{n} \quad \text{(eq. 6)}$$

where $n$ is the number of segments (i.e. amino acids) and $a$ is a scaling parameter related to the distance between two segments. Using distance distribution between adjacent FSFG motifs obtained from our previous MD simulation (Raveh et al., 2016) as $L = \langle r^2 \rangle^{1/2} \approx \langle r \rangle = 33.0$ Å for adjacent FSFG repeats (where $n = 15$), $a$ was determined as 8.5 Å. The $\langle r^2 \rangle^{\frac{1}{2}}$ for $i$, $i+2$ FSFG motifs ($n = 34$) would then be 49.7 Å. Though this estimate does not account for increased size of FSFG-K upon TF interaction as observed here, the estimated $\langle r^2 \rangle^{\frac{1}{2}}$ is similar to the peak in the $P(r)$ at 52 Å for [$^2$H]-FSFG-K in the presence of both NTF2 and Kap95.

Ensemble analysis

Ensemble analysis was performed using the EOM 2.0 software from the ATSAS package (Tria et al., 2015). This approach involved generating a pool of random coil protein models. This was performed using trajectory directed ensemble sampling traDES (Feldman and Hogue, 2000, 2002) to produce a pool of 100,000 possible structures. Theoretical scattering of each model was then computed using CRYSON from the ATSAS package (Svergun et al., 1998). This was performed twice, calculating for 42% D$_2$O and 80% deuteration of [$^2$H]-FSFG and calculating for 92% D$_2$O and 0% deuteration. The genetic algorithm method, GAJOE, within EOM 2.0 package was subsequently used to select from the pool of theoretical scattering curves, a subset of structures whose weighted average scattering curve, that best fits the contrast matching data. GAJOE was run with default parameters except for allowing the algorithm to use a minimum of 1 curve per ensemble (default = 5) and not enabling the subtraction of a constant value. GAJOE was repeated 1,000 times and no constant subtraction was used.

Generation of a pool of conformers from ANTON simulations

All-atom molecular dynamics simulations of FSFG-K with NTF2 using the TIP4pD water model (Piana et al., 2015) was performed previously using the Anton supercomputer (Raveh et al., 2016). The trajectory consisted of 19,930 frames with a frame rate of 0.06 ns/frame, totaling ~1.2 μs. The trajectory was loaded into VMD (Humphrey et al., 1996) and PDB files of FSFG-K were written for each frame. As above, CRYSON was used to compute the theoretical scattering for each model and GAJOE was used for ensemble selection.

---

**Supplementary Item titles.**

**Figure S1, related to Figure 1: SANS data analysis obtained from the dilution series of free FSFG-K and free [$^2$H]-FSFG-K.**

**Figure S2, related to Figure 2: Contrast match and inverse contrast match SANS data of FSFG-K in the presence of transport factors as well as dynamic light scattering data of similarly prepared samples.**

**Figure S3, related to Figure3: Averaged and individual Cα-Cα distance maps computed from the conformers that comprise the selected ensemble based on a selection from the "random pool."**

**Figure S4, related to Figure3: Validation of ensemble analysis from the contrast match SANS of [$^2$H]-FSFG-K in the absence and presence of transport factors.**

**Table S1: related to STAR Methods section Worm-like chain modeling: Best fitted parameters determined by fitting the experimental data to the Worm-like chain model.**

**Table S2, related to Figure 3: Results of the EOM fitting of the experimental data.**

---

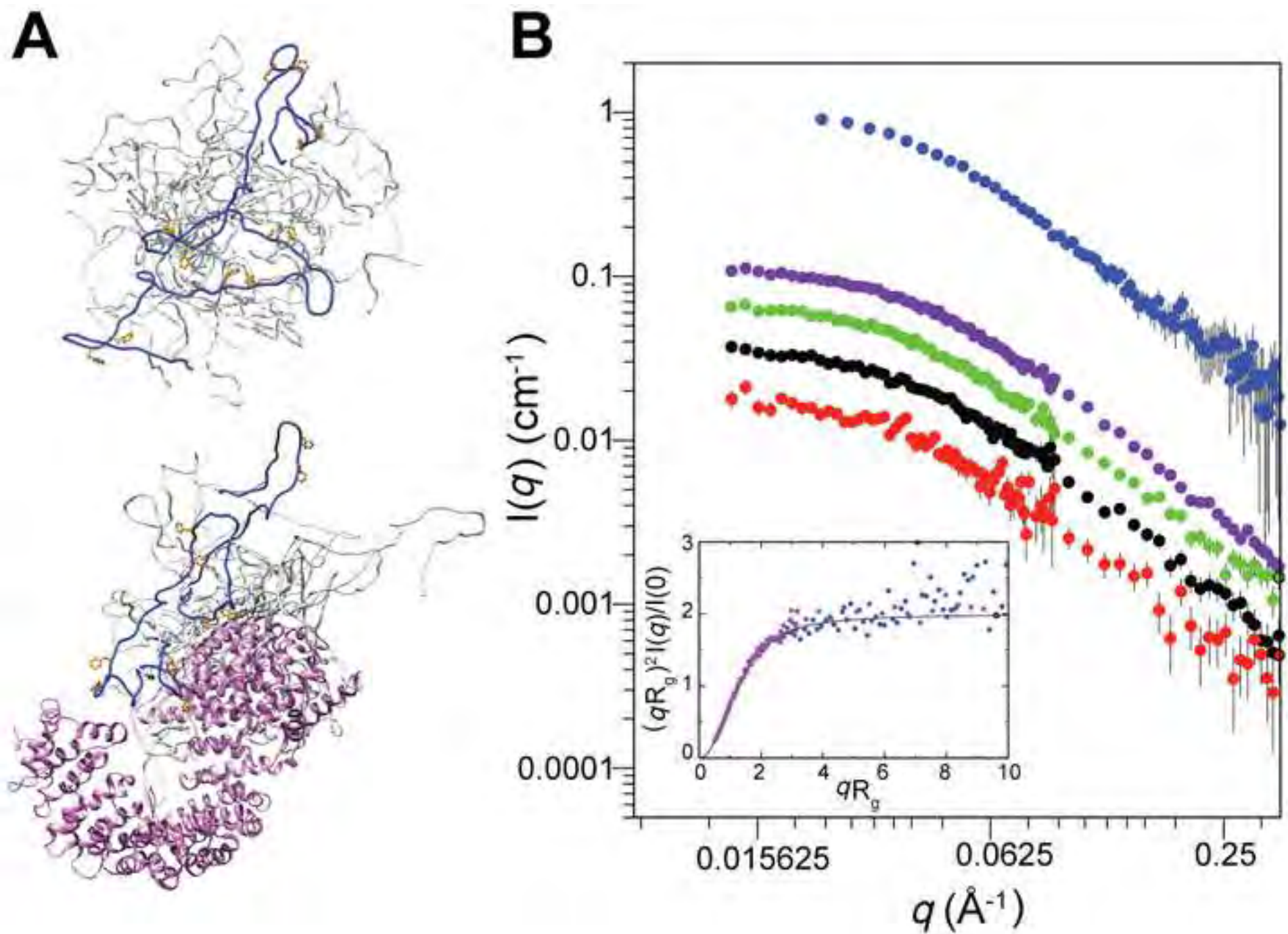

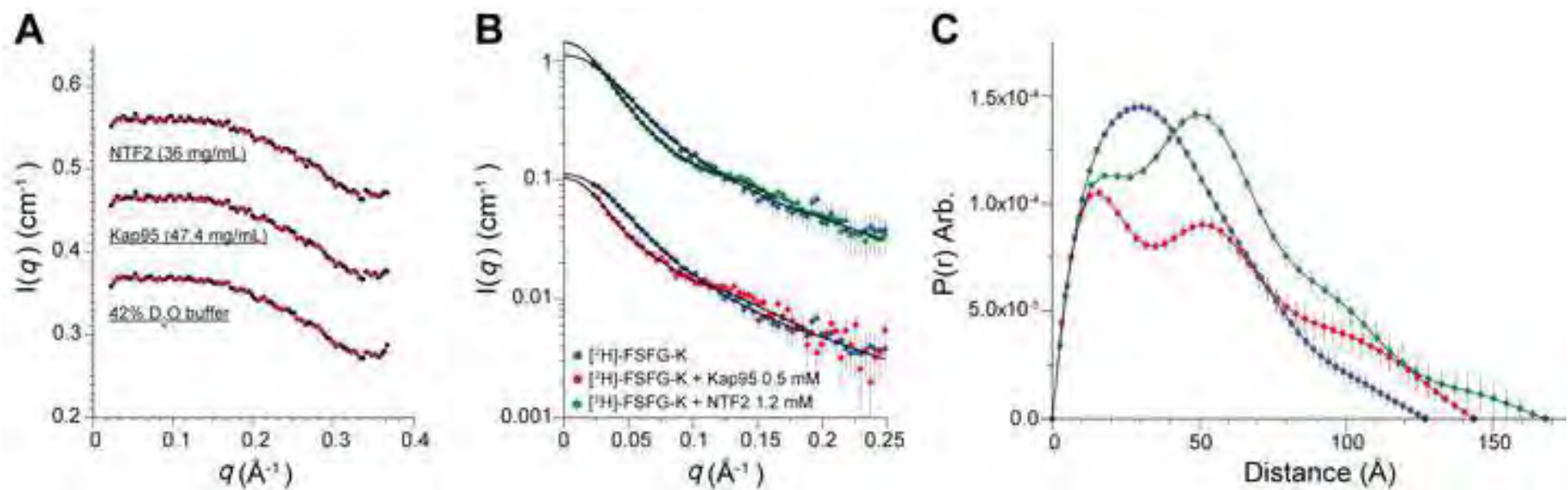

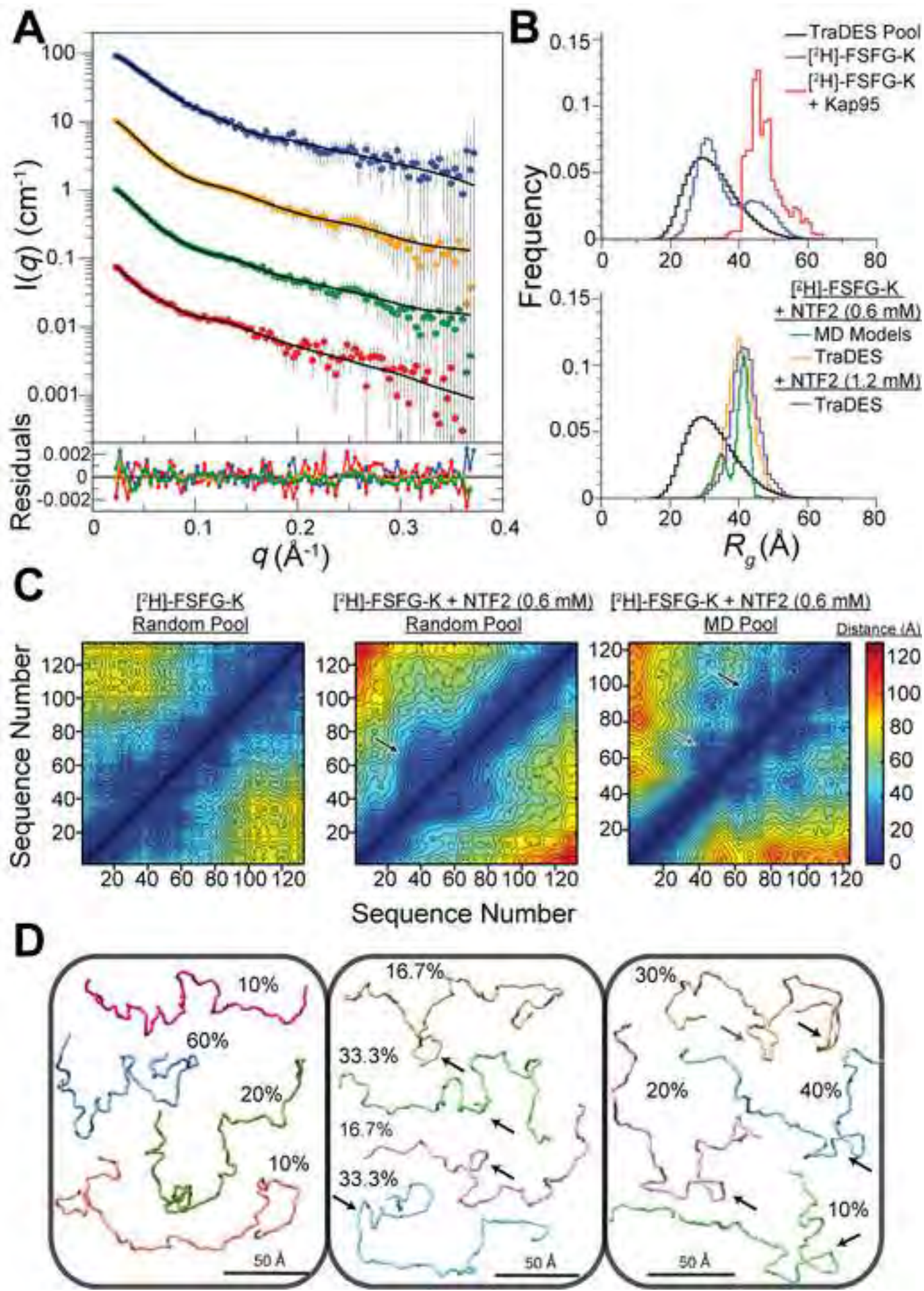

**KEY RESOURCES TABLE**

| REAGENT or RESOURCE | SOURCE | IDENTIFIER |
|---|---|---|
| Bacterial and Virus Strains | | |
| BL21-Gold(DE3) strain | Agilent Technologies | #230132 |
| Chemicals, Peptides, and Recombinant Proteins | | |
| Deuterium oxide | Cambridge Isotope Lab. | DLM-6-1000 |
| Glutathione sepharose 4B | GE Health | Cat#17075601 |
| TALON resin | Clonetech | Cat#635501 |
| Human alpha thrombin | Enzyme research laboratories | Cat#HT1002a |
| Size exclusion column, Superdex S200 | GE Health | 50005 |
| **Recombinant DNA** | | |
| Plasmid: pET-FSFG-K-His | (Hough, 2015) | N/A |
| Plasmid pGEX-GST-Kap95 | (Hough, 2015) | N/A |
| Plasmid: pRSF-NTF2-His | This paper | N/A |
| **Software and Algorithms** | | |
| Ensemble from simulation | (Raveh, 2016) | https://www.ncbi.nlm.nih.gov/pubmed/27091992 |
| PDB generation, VMD | U. Illinois | http://www.ks.uiuc.edu/Research/vmd/ |
| SCATTER | BioIsis | http://www.bioisis.net/tutorial/9 |
| DYNAMICS 7.1.0.25 | Wyatt Technology | https://www.wyatt.com/products/software/dynamics.html |
| **Other** | | |
| SANS measurements | Bio-SANS Oak Ridge | http://neutrons.ornl.gov/biosans |

**Supplemental Information**

Deciphering the 'fuzzy' interaction of FG nucleoporins and transport factors using SANS

Samuel Sparks, Deniz B. Temel, Michael P. Rout, and David Cowburn

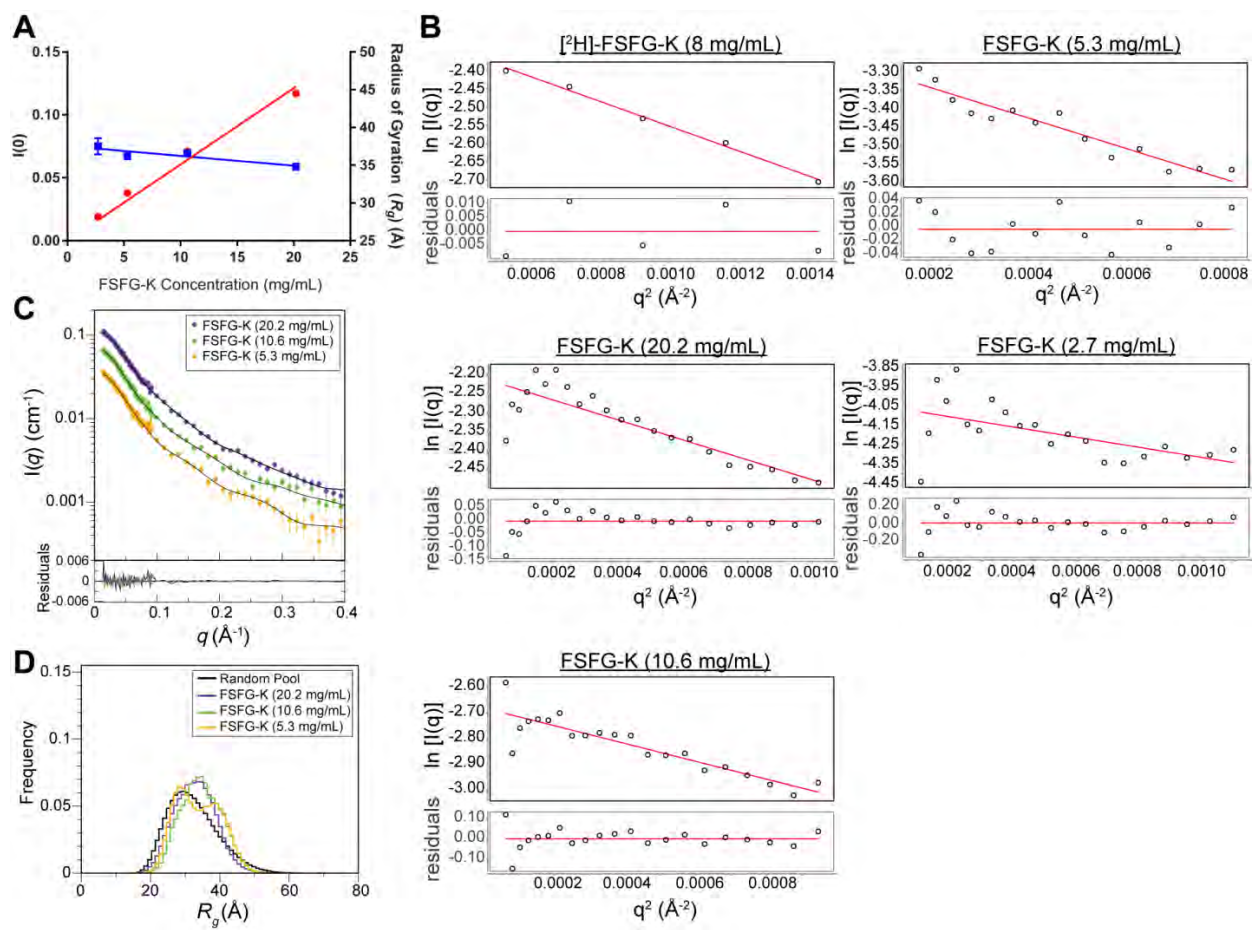

**Figure S1, related to Figure 1: SANS data analysis obtained from the dilution series of free FSFG-K and free [²H]-FSFG-K.** (A) Concentration dependence of the forward scattering, $I_0$, (red) and the radius of gyration, $R_g$, (blue) from SANS samples of natural abundance FSFG-K in 92% $D_2O$. Solid lines are the fits to a linear regression.

(B) Guinier plots and plots of the residuals of the Guinier fit of the FSFG-K dilution in 92% $D_2O$ and free form of [²H]-FSFG-K in 42% $D_2O$. Guinier fits were performed within the limit of $q_{max}R_g$ < ~1.0 (See STAR Methods section). See Table 1 for calculated $R_g$ and $I_0$ values, as well the $q_{max}R_g$ limit used. The absence of concentration-dependent effects in $I_0$ and $R_g$ in (A) and linear Guinier plots shown in (B) and indicates these samples are monodisperse.

(C) EOM fit from the selected ensemble (solid line) to the experimental data for samples of different concentrations of FSFG-K in 92% $D_2O$ buffer. The "random pool" was used for the selection. The plot of the residuals of the fit is displayed in the panel below.

(D) Comparison of the frequency distribution of $R_g$s from the optimized ensembles of FSFG-K at indicated concentrations to the distribution of $R_g$ from "random pool" models (black).

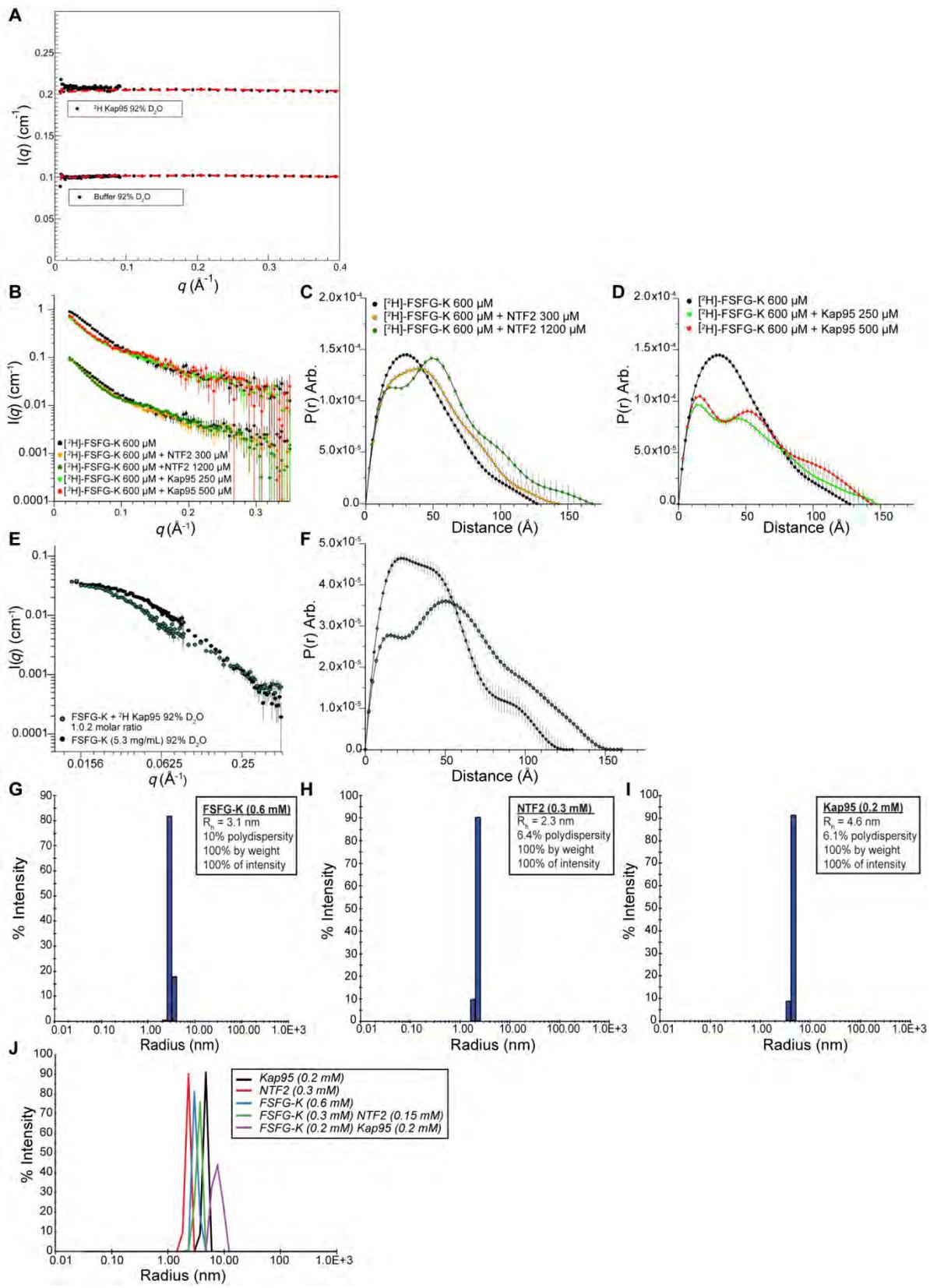

**Figure S2, related to Figure 2: Contrast match and inverse contrast match SANS data of FSFG-K in the presence of transport factors as well as dynamic light scattering data of similarly prepared samples.** (A) Quality of inverse contrast match of partially deuterated (62% deuteration of non-exchangeable H's) Kap95 in 92% $D_2O$ buffer. The black dots are the SANS data points of the partially deuterated Kap95 (100 μM) (top) and the 92% $D_2O$ buffer (bottom). The dashed red lines are the moving average of the 92% $D_2O$ buffer-only scattering. Data is vertically displaced for clarity and truncated to $q = 0.4$ Å$^{-1}$.

(B-D) Comparison of SANS data of [$^2$H]-FSFG-K acquired in the presence of higher transport factor concentrations (Kap95 (0.5 mM), red, and NTF2 (1.2 mM), dark green) with data acquired at lower TF concentration (Kap95 (0.25 mM), light green, and NTF2 (0.3 mM), orange). Features in the scattering profile (B) at $q = 0.12$ Å$^{-1}$ are less pronounced at a lower molar ratio which is reflected by a reduced or shifted distribution in the population of inter-nuclear distances at ~50 Å computed by the radial distribution function in the presence of (C) NTF2 and (D) Kap95. (E) Log-log plot from the inverse contrast matching SANS of the natural abundance FSFG-K (375 μM) with (green) and without (black) partially [$^2$H]-Kap95 (75 μM) at 92% $D_2O$.

(F) Radial distribution function computed from the data in (E).

(G-J) Intensity-weighted DLS of free forms of (G) FSFG-K, (H) NTF2, (I) Kap95. For (G-I) the concentration of the sample, radius of hydration, $R_h$, polydispersity, and percentage of the main peak weighted by mass and intensity are indicated. (J) DLS from samples of FSFG-K in the presence of either Kap95 or NTF2 at indicated concentrations compared to the individual components in their free forms. The $R_h$ for FSFG-K (0.3 mM) and NTF2 (0.15 mM) was 3.7 nm (21.4% polydispersity) while the sample containing FSFG-K (0.2 mM) and Ka95 (0.2 mM) had an $R_h$ of 7.6 nm (17.8 % polydispersity).

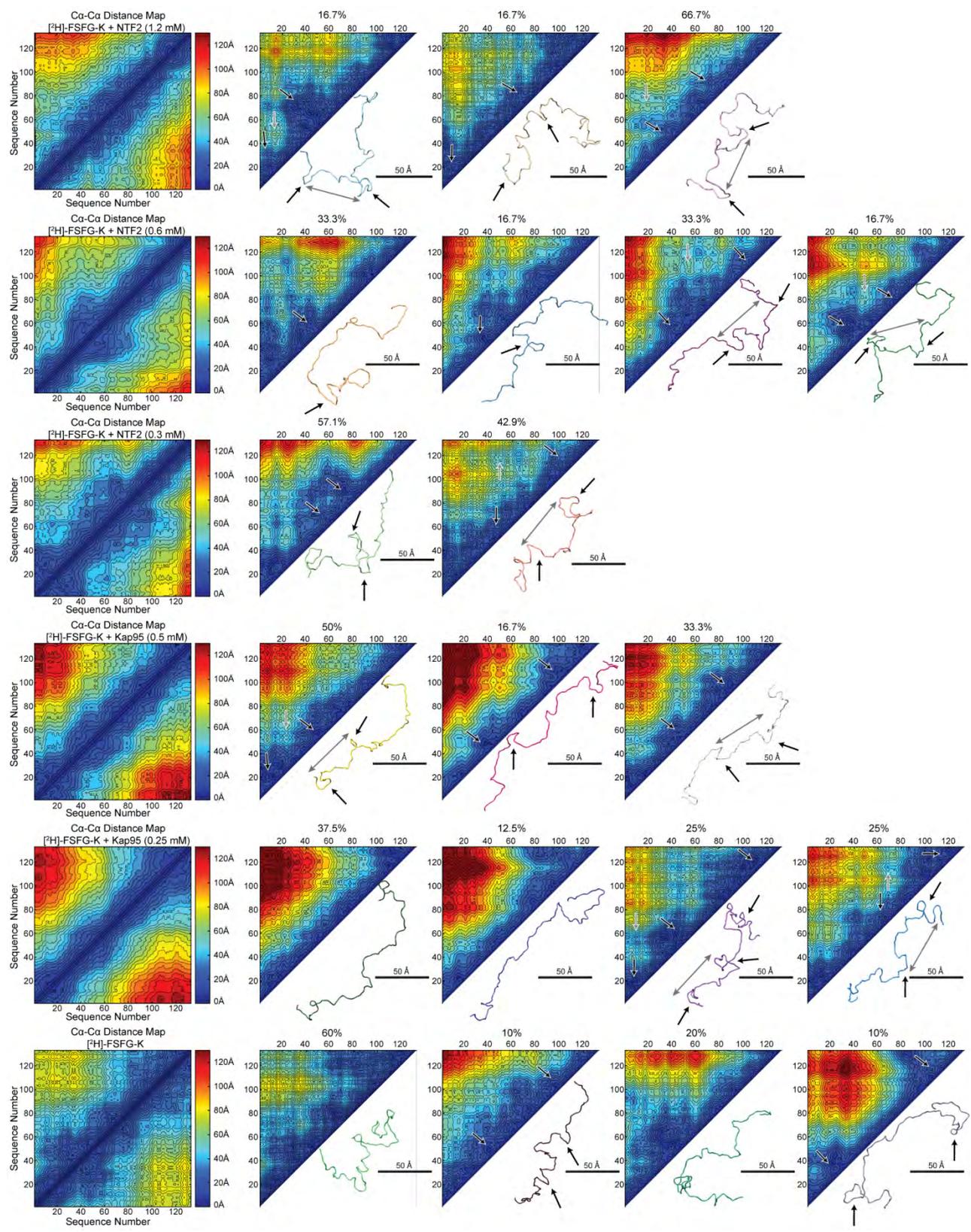

**Figure S3, related to Figure3: Averaged and individual Cα-Cα distance maps computed from the conformers that comprise the selected ensemble based on a selection from the "random pool."** The leftmost column plots the weighted average Cα-Cα distance map based on EOM fitting to the SANS data from the indicated samples. The Cα-Cα distance map and the backbone structure from each individual conformer are shown as well as the relative fraction that each conformer contributed to the final selected ensemble. The structures of the individual conformers are orientated such that the N-terminal residue or C-terminal residue points towards the bottom left or top right of the panel, respectively. The $\chi^2$, selected ensemble $R_g$ and $D_{max}$ and the end-to-end (Cα(N)-Cα(C)) distance for each ensemble are given in Table S2. When observed, a black arrow pointing to the selected conformers indicates local clustering in the form of a loop or turn-like feature. On the individual distance maps, the black arrows indicate the sequence position of the observed loop or turn feature. Grey arrows on the distance maps indicate Cα-Cα distances that are ~50 Å from two loop features and the grey double-sided arrows (equivalent to 50 Å) approximates this on the structure of the selected ensembles.

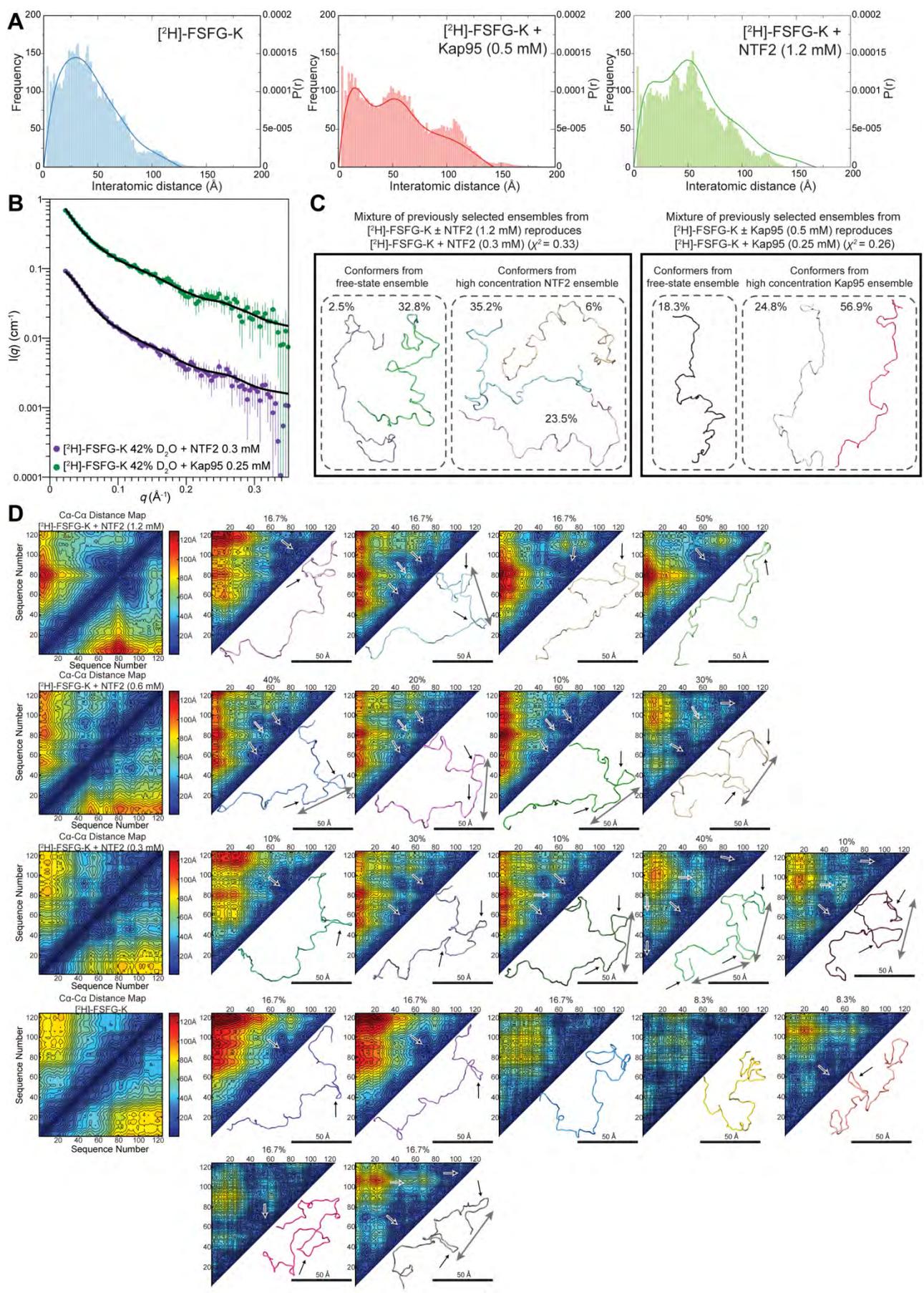

**Figure S4, related to Figure3: Validation of ensemble analysis from the contrast match SANS of [²H]-FSFG-K in the absence and presence of transport factors.** (A) Comparison of the radial distribution function (solid lines, data from Figure 2C) with the weighted average histogram of each Cα-Cα distance in each conformer that comprises the selected ensembles from the data sets indicated. The histograms were plotted as 1 Å bins ranging from 1-150 Å for free [²H]-FSFG-K and [²H]-FSFG-K with NTF2 (1.2 mM) and from 1-200 Å [²H]-FSFG-K with Kap95 (0.5 mM).

(B-C) Multicomponent mixture analysis of the selected ensembles from (the sample of free [²H]-FSFG-K and of [²H]-FSFG-K in the presence of the highest concentrations of transport factor measured. (B) The software OLIGOMER from the ATSAS package was used to fit the experimental SANS curves of the [²H]-FSFG-K in the presence of the lowest concentrations of transport factor measured (NTF2 0.3 mM and Kap95 0.25 mM) with the theoretical scattering from some mixture of the selected conformers previously optimized by EOM for free [²H]-FSFG-K and [²H]-FSFG-K with the highest concentration of transport factor (NTF2 1.2 mM and Kap95 0.5 mM). The fit to the scattering profile is shown as a solid black line. The scattering data is vertically translated for clarity. (C) The conformers within the new ensemble for each transport factor which contains mixtures of both free [²H]-FSFG-K and [²H]-FSFG-K with highest concentration transport factor whose weighted average theoretical scattering profile fits the data with the $\chi^2$ values indicated.

(D) Averaged and individual Cα-Cα distance maps computed from the conformers that comprise the selected ensemble based on a selection from the "MD pool." The leftmost column plots the weighted average Cα-Cα distance map based on EOM fitting to the SANS data from the indicated samples. The Cα-Cα distance map and the backbone structure from each individual conformer are shown as well as the relative fraction that each conformer contributed to the final selected ensemble. The structures of the individual conformers are orientated such that the N-terminal residue or C-terminal residue points towards the bottom left or top right of the panel, respectively. The $\chi^2$, selected ensemble $R_g$ and $D_{max}$ and the end-to-end (Cα(N)-Cα(C)) distance for each ensemble are given in Table S2. When observed, a black arrow pointing to the selected conformers indicates local clustering in the form of a loop or turn-like feature. On the individual distance maps, the black arrows indicate the sequence position of the observed loop or turn feature. Grey arrows on the distance maps indicate Cα-Cα distances that are ~50 Å from two loop features and the grey double-sided arrows (equivalent to 50 Å) approximates this on the structure of the selected ensembles.

# Supplementary Tables

| Sample | $b$ (Å) | $np$ | $L$ (Å) | $R_g$ (Å) |
|---|---|---|---|---|
| [²H]-FSFG-K [0.6 mM] (8mg/ml) | 17.8 ± 4.5 | 2.4 | 477.6* | 36.6 |
| [²H]-FSFG-K [0.6 mM] Kap95 [0.25 mM] | 60.7 ± 1.6 | 8.0 | 236.1 ± 3.3 | 40.9 |
| [²H]-FSFG-K [0.6 mM] Kap95 [0.5 mM] | 67.1 ± 2.8 | 8.9 | 247.8 ± 6.3 | 43.7 |
| [²H]-FSFG-K [0.6 mM] NTF2 [0.3 mM] | 25.1 ± 2.1 | 3.3 | 477.6* | 43.0 |
| [²H]-FSFG-K [0.6 mM] NTF2 [0.6 mM] | 29.4 ± 2.7 | 3.9 | 477.6* | 46.2 |
| [²H]-FSFG-K [0.6 mM] NTF2 [1.2 mM] | 33.4 ± 2.4 | 4.4 | 472.3 ± 23.4 | 48.6 |
| FSFG-K [375 μM] + [²H]-Kap95 [75 μM] | 48.6 ± 7.6 | 6.4 | 423.4 ± 48.6 | 53.9 |
| FSFG-K (20.2 mg/mL) | 24.3 ± 6.1 | 3.2 | 333.3 ± 66.8 | 34.8 |
| FSFG-K (10.6 mg/mL) | 20.1 ± 9.1 | 2.7 | 426.7 ± 166.9 | 36.5 |
| FSFG-K (5.3 mg/mL) | 31.7 ± 8.3 | 4.2 | 289.5 ± 54.2 | 36.1 |
| FSFG-K (2.7 mg/mL) | 33.7 ± 20.5 | 4.4 | 293.9 ± 125.5 | 37.4 |

**Table S1: related to STAR Methods section Worm-like chain modeling: Best fitted parameters determined by fitting the experimental data to the Worm-like chain model.** (*) indicates the parameter was constrained to its theoretical limit (see STAR Methods section).

| Sample | $\chi^2$ | Selected ensemble $R_g$ (Å) | Selected ensemble $D_{max}$ (Å) | Selected ensemble end-to-end distance (Å) Cα(N)-Cα(C) |
|---|---|---|---|---|
| **Selected ensembles when "Random pool" was used for selection** | | | | |
| [$^2$H]-FSFG-K [0.6 mM] (8mg/ml) | 0.20 | 35.5 | 87.4 | 73.7 |
| [$^2$H]-FSFG-K [0.6 mM] Kap95 [0.25 mM] | 0.26 | 46.9 | 129.1 | 118.0 |
| [$^2$H]-FSFG-K [0.6 mM] Kap95 [0.5 mM] | 0.38 | 47.0 | 137.4 | 135.0 |
| [$^2$H]-FSFG-K [0.6 mM] NTF2 [0.3 mM] | 0.17 | 38.5 | 119.0 | 119.0 |
| [$^2$H]-FSFG-K [0.6 mM] NTF2 [0.6 mM] | 0.15 | 40.7 | 117.6 | 115.3 |
| [$^2$H]-FSFG-K [0.6 mM] NTF2 [1.2 mM] | 0.27 | 42.0 | 141.0 | 128.1 |
| FSFG-K [375 mM] + [$^2$H]-Kap95 [75 μM] | 0.99 | 46.6 | 120.9 | 108.3 |
| FSFG-K (20.2 mg/mL) | 0.83 | 33.3 | 80.0 | 62.7 |
| FSFG-K (10.6 mg/mL) | 0.54 | 35.4 | 86.8 | 72.1 |
| FSFG-K (5.3 mg/mL) | 0.51 | 34.9 | 91.3 | 82.1 |
| FSFG-K (2.7 mg/mL) | 0.93 | 35.4 | 89.6 | 75.5 |
| **Selected ensembles when "MD pool" was used for selection** | | | | |
| [$^2$H]-FSFG-K [0.6 mM] (8mg/ml) | 0.22 | 37.6 | 92.0 | 89.8 |
| [$^2$H]-FSFG-K [0.6 mM] NTF2 [0.3 mM] | 0.20 | 37.5 | 91.5 | 86.9 |
| [$^2$H]-FSFG-K [0.6 mM] NTF2 [0.6 mM] | 0.19 | 39.6 | 103.5 | 99.6 |
| [$^2$H]-FSFG-K [0.6 mM] NTF2 [1.2 mM] | 0.37 | 41.2 | 125.0 | 90.1 |

**Table S2, related to Figure 3: Results of the EOM fitting of the experimental data.**